\documentclass[twocolumn,trackchanges]{aastex631}

\usepackage{amsmath}
\usepackage{ulem}

\newcommand{\fuernkranz}{F\"urnkranz}

\newcommand\ltau{\ensuremath{{\mathrm{log}\:\tau}}}

\graphicspath{{./}{figures/}}

\begin{document}


\title{The Age Distribution of Stellar Orbit Space Clumps}


\correspondingauthor{Verena \fuernkranz}
\email{fuernkranz@mpia.de}

\author[0000-0002-9460-0212]{Verena \fuernkranz}
\affiliation{Max-Planck-Institut f\"ur Astronomie \\
K\"onigstuhl 17, D-69117\\
Heidelberg, Germany}

\author[0000-0003-4996-9069]{Hans-Walter Rix}
\affiliation{Max-Planck-Institut f\"ur Astronomie \\
K\"onigstuhl 17, D-69117\\
Heidelberg, Germany}

\author[0000-0001-6629-0731]{Johanna Coronado}
\affiliation{Max-Planck-Institut f\"ur Astronomie \\
K\"onigstuhl 17, D-69117\\
Heidelberg, Germany}

\author[0000-0001-8898-9463]{Rhys Seeburger}
\affiliation{Max-Planck-Institut f\"ur Astronomie \\
K\"onigstuhl 17, D-69117\\
Heidelberg, Germany}


\begin{abstract}

The orbit distribution of young stars in the Galactic disk is highly structured, from well-defined clusters to streams of stars that may be widely dispersed across the sky, but are compact in orbital action-angle space. The age distribution of such groups can constrain the timescales over which co-natal groups of stars disperse into the `field'. Gaia data have proven powerful to identify such groups in action-angle space, but the resulting member samples are often too small and have too narrow a CMD coverage to allow robust age determinations. Here, we develop and illustrate a new approach that can estimate robust stellar population ages for such groups of stars. This first entails projecting the predetermined action-angle distribution into the 5D space of positions, parallaxes and proper motions, where much larger samples of likely members can be identified over a much wider range of the CMD. It then entails isochrone fitting that accounts for a) widely varying distances and reddenings; b) outliers and binaries;  c) sparsely populated main sequence turn-offs, by incorporating the age information of the low-mass main sequence; and d) the possible presence of an intrinsic age spread in the stellar population.  When we apply this approach to 92 nearby stellar groups identified in 6D orbit space, we find that they are predominately young ($\lesssim 1$~Gyr), mono-age populations. Many groups are established (known) localized clusters with possible tidal tails, others tend to be widely dispersed and manifestly unbound. This new age-dating tool offers a stringent approach to understanding on which orbits stars form in the solar neighborhood and how quickly they disperse into the field.

\end{abstract}


\keywords{Milky Way dynamics (1051) --- Milky Way disk (1050) --- Solar neighborhood(1509) --- Stellar kinematics (1608) --- Star clusters (1567) --- Stellar associations (1582) --- Galactic archaeology(2178)}


\section{Introduction} \label{sec:intro}

Stars are born in a highly clustered fashion, in position, velocity, and orbit space \citep[e.g.][]{lada2003,Lamb2010,Kounkel2019}. It appears that the vast majority of them do not remain in bound clusters after the gas expulsion phase \citep[e.g.][]{Chandar2017}. And even those that first remain in ``open'' clusters beyond this phase tend to eventually become unbound from these clusters, by internal two- or three-body interactions, or by external tides. Consequently, most stars with ages $>$ 5~Gyrs have become ``field stars''. However, how clustered (bound or unbound) the full population of stars was in their first, say, 10~Myrs, and how, or how rapidly, the transition from star clusters and associations to the Galactic field takes place has yet to be fully understood. Studying this transition process will open a new window on constraining whether all stars are born in clusters, give insights into how hierarchical star formation is, and allow one to tackle how rapidly the dissolution of clusters takes place.

The Gaia mission has provided data that are revolutionary for our understanding of the formation and structure of the stellar component of the Milky Way disk \citep{gaia2022}. The third Gaia data release (Gaia DR3) on June 13, 2022 has provided precise 6D position measurements ($\alpha, \delta, \varpi$), proper motion ($\mu_{\alpha}, \mu_{\delta}$), and radial velocity measurements ($v_{r}$) for millions of stars in the extended solar neighborhood \citep{gaiadr32022}. Recent work, e.g. by \citet{trick2019} and \citet{coronado2020}, had shown that this 6D parameter space -- or equivalently \emph{orbit space} -- is highly clustered over a wide range of scales. Large-scale structures might be caused by resonances of the Galactic bar and spiral arms, or by satellite infall into the Milky Way Galaxy \citep{katz_2018}. Small-scale clustering represents the concepts of open clusters, associations, and stellar (disk) streams. 

In the immediate neighborhood, the space velocities can serve as orbit proxies. On larger scales, action-angle coordinates (J, $\Theta$) stand out as a powerful coordinate system to search for substructure in the Galactic orbit and orbital phase space: the three actions denote the ``orbit''; the three angles the ``orbital phase``. Recent work by \citet{coronado2022} has shown that we can find numerous stellar groups by linking stars in this 6D parameter space: some are simply clusters, others extend over 100's of parsecs. In \citet{coronado2022} we discovered that the same orbits that host one of these groups very frequently host numerous other distinct orbit-phase overdensities along these same orbits, which we call \emph{ pearls}\footnote{These pearls are strung along one orbit, like pearls on a string.}. This might open up a new way to probe the filamentary structure of the ISM.

The central overall goal of our present work is to associate \emph{ages} to stellar action-angle groups in the Galactic disk. These are crucial data for understanding for how long ensembles of stars remember their initially near-identical orbits and orbital phases. Once sets of stars (clusters or associations) are no longer gravitationally bound but still tightly clustered in orbit space (J, $\Theta$), they will continue to disperse into the field by simple phase-mixing and by further relative orbit diffusion, driven e.g. by the Galactic tidal field. To make predictions about the rate of this dispersal, age information about the ensembles of stars is needed. 

For coeval ensembles of stars, fitting isochrones to their CMD positions is a long-established approach to age estimates. This suggests that one should simply fit the CMDs of the stellar ensembles identified \citep[see][]{coronado2022} in 6D phase space. However, radial velocities from Gaia exist only for stars over a limited range of effective temperatures (3.100 to 14.500~K) and for stars brighter than $G_{RVS}\approx 14$~mag. Furthermore, the radial velocities for hot and cool very young stars are not very precise ($\delta v_{r}\approx 5$~km~s$^{-1}$), which is a serious impediment in orbit clustering studies. At any rate, this means that the CMDs of these 6D sample are incomplete or missing for the uppermost main sequence (including potentially the turn-off for ages $\tau<200$~Myrs) and  for the lowest-mass stars, which are also good age diagnostics as stars may take $>10^8$~years to settle on the zero-age main sequence.  

The practical goal of this paper is to overcome this limitation by projecting the tight 6D action-angle distribution of these stellar ensembles to 5D position-proper motion space, where we can then establish much larger membership samples with higher completeness at the extreme mass and temperature ranges that are so precious for age estimates. This is possible since Gaia is highly complete to $G\approx 19$~mag across a very broad range of colors, if only positions, parallaxes, and proper motions are required.  We aim to end up with CMD's that are much better populated in both total sample size and color range. This requires that the groups, high-density peaks in 6D (J, $\Theta$), still have a distinct density contrast in their 5D projection ($\alpha, \delta, \varpi, \mu_{\alpha}, \mu_{\delta}$, but without $v_{r}$).

Conceptually, our approach to extending group membership is as follows: we start with a set of stellar groups identified in 6D space. Specifically, we get these by identifying 6D action-angle groups within 800~pc from the Sun in a similar fashion to \cite{coronado2022}. We use the DBSCAN clustering algorithm and choose a search setting that identifies 102 action-angle groups of stars, which is about twice the sample size as the 55 action-angle groups identified in \cite{coronado2022}. For our analysis, we pick all groups with at least 20 member stars. This reduced set of 94 groups largely overlaps with the open cluster catalog published in \citet{cantatgaudin_2020}, as well as with the stellar groups detected in \citet{hunt_2023}. Additionally, almost all of our stellar groups can be found in the catalog of \citet{kounkel_2019}, which is known to be complete but highly contaminated (see \citet{zucker_2022}). We project these groups into the 5D position-proper motion space, where we have a factor $\sim$8 more stars in the Gaia catalog. For each action-angle group, we assign additional candidate members that are located in the same 5D region. Specifically, we run a kernel density estimate on the 5D distribution of each group and pick all stars in the Gaia catalog that show an excess at high densities, compared to stars in the Gaia mock catalog \citep{rybizki_2020}. 

We then take 92 of these groups with their extended membership to develop, test and apply our new approach to age dating these groups of potentially co-natal stars, accounting for their potentially sparse and widely dispersed nature. This approach entails likelihood isochrone fitting that takes into account a) differential reddening, b) the widely ranging member distances, c) the individual photometry and distance errors, as well as d) binaries and outliers. We also test whether groups can be described as \emph{effectively mono-age}, rather than have evidence for a large age dispersion in their CMD. Accounting for all these aspects in age-dating is crucial in this regime, and has not been implemented before.

With this new approach, we determine the ages and possible intrinsic age spreads of the identified stellar action-angle groups, and find them to be predominately young ($\tau \lesssim 2\, \tau_{orbit}$) and mono-age.  The analysis of the dispersal of these groups in orbit and orbital phase-space will be subject of upcoming work, as will be the age distribution of any ``pearls'' along the same orbit, as well as the smallest and most diffuse groups we can identify that still show mono-age CMD sequences.

The paper is organized as follows. In Section~\ref{sec:data} we describe the data used in this study. Section~\ref{sec:groups} outlines our method for finding groups in 6D action-angle space and associating member stars by  extending our search to the 5D position proper motion space. In Section~\ref{sec:ages} we present our likelihood isochrone fitting method which determines ages of the groups. Our results are discussed in Section~\ref{sec:results}. Finally, we summarize our findings and conclude in Section~\ref{sec:summary}.

\section{Data} \label{sec:data}

\subsection{Excerpts of the Gaia DR3 catalog} \label{gaia}

We base our analysis exclusively on Gaia DR3 \citep{gaiacollab_2022}. As our intention is to study the extended solar neighborhood, we limit the analysis to stars with parallax $\geq$ 1~mas, corresponding to a maximum distance of 1 kpc from the Sun. We clean this sample from sources with spurious astrometry, following \citet{rybizki_2022}, and exclude sources with fidelity\_v2 $<$~0.9. We also limit the analysis to stars with parallax\_over\_error $\ge$~10, to eliminate distance uncertainties as the main source of orbit uncertainty. This results in a database containing 41.138.028~sources, and we will refer to this catalog as the \textit{Gaia sample} throughout.

As we start our analysis by identifying groups in the 6D action-angle space, we also introduce a slightly more confined sample with $\varpi \ge $~1.25 (within 800~pc from the Sun), comprising only the stars that also have a radial velocity ($v_{r}$) measurement. We adopt the same fidelity cut, but relax it to parallax\_over\_error $\ge$~3 to get a larger 6D sample. This 6D database contains 9.080.884~sources, and we will refer to it as \textit{Gaia RVS sample}.

As we need to fit isochrones to dereddened colors and magnitudes, we query extinction values (A$_{G}$, A$_{BP}$, A$_{RP}$) from the astrophysical parameter catalog produced by the Apsis processing chain developed by the Gaia DPAC \citep{fouesneau_2022}. Specifically, we query all stars with an extinction entry available, and within 1~kpc from the Sun. Additionally, we limit the catalog to zero-age main sequence stars and apply a 2~mag~$<$~M$_G$~$<$ 8~mag cut (see Section~\ref{reddening}), which reduces the catalog to 23.543.432~sources and hereafter we will refer to it as \textit{Gaia extinction sample}. For the most part, it encompasses the \textit{Gaia sample}. 

\subsection{Gaia EDR3 mock catalog} \label{mock}

Since we will identify overdensities in 5D position-proper motion space, we will need to know what densities we should expect from a smooth distribution.
To this end, we make use of the Gaia EDR3 mock catalog by \citet{rybizki_2020}, which presents an analogous dataset to the Gaia DR3 catalog, but with a smooth and fully phase-mixed orbit distribution. We apply the same distance cut by only selecting stars with $\varpi \geq $~1, and add {\tt WHERE popid != 11} into the ADQL query, which removes clusters and moving groups. These are otherwise included in the catalog by default. The resulting dataset comprises 46.830.045~stars, and will be named \textit{Gaia mock sample} throughout the manuscript. The size of the catalog is comparable to the size of the \textit{Gaia sample} catalog, and therefore it is well suited for comparison.

\subsection{PARSEC isochrones} \label{subsec:iso}

For age determinations via isochrone fitting, we need to choose one (or more) isochrone models. After considerable experimentation, we settled on PARSEC isochrones \citep{bressan_2012} for our analysis, as they provide the most consistent age estimates between the turn-off and the low-mass main sequence. MIST isochrones \citep{dotter2016,choi2016} provide less consistent age estimates for our set of action-angle groups, as they are not optimized for the lower main sequence.

Specifically, we download the default isochrone tables for a set of 6 metallicities in the range of [Fe/H] $= -0.4$ to $+0.1$, and 107~ages from log $\tau = 5$ to $10.3$, resulting in an isochrone table comprising 224.951~entries. 

Our analysis requires that each individual isochrone can return finely spaced magnitude and color values, which we ensure by linear interpolation. Specifically, for each individual isochrone we loop through all mass points and interpolate Mass, G$_{mag}$, BP$_{mag}$, and RP$_{mag}$ whenever $\Delta$G$_{mag} >$~0.03 mag between two mass points. After applying this method, we doubled the isochrone entries to a total of 458.181. We will refer to this setup as \textit{isochrone table} in the text. 

\section{Stellar Groups in Action-Angle Space } \label{sec:groups}

We now describe membership identification of the stellar groups whose ages are at the heart of our analysis: their initial identification as prominent overdensities in the 6D action-angle space and the subsequent extended membership identification after projection of each action-angle distribution ("patch") into the 5D position-proper motion space.

\subsection{Action-angle computation} \label{subsec:actionangles}

The calculation of actions (J) and angles ($\Theta$) requires both the full 6D position-velocity information in Gaia's original coordinates ($\alpha, \delta, \varpi, \mu_{\alpha}, \mu_{\delta}, v_{r}$) and an assumed gravitational potential. For the potential, we adopt Galpy's axisymmetric MWPotential2014 \citep{bovy_2015}, which has adopted a circular velocity of 220~km~s$^{-1}$ at the Solar radius of 8~kpc, and which comprises a bulge, a disk, and a dark-matter component \citep{bovy_2015}. We then use the \emph{St\"ackel Fudge} algorithm by \citet{binney_2012} to estimate the actions and angles for all stars in our \textit{Gaia RVS sample}.

\begin{figure*}[ht!]
\includegraphics[width=\textwidth]{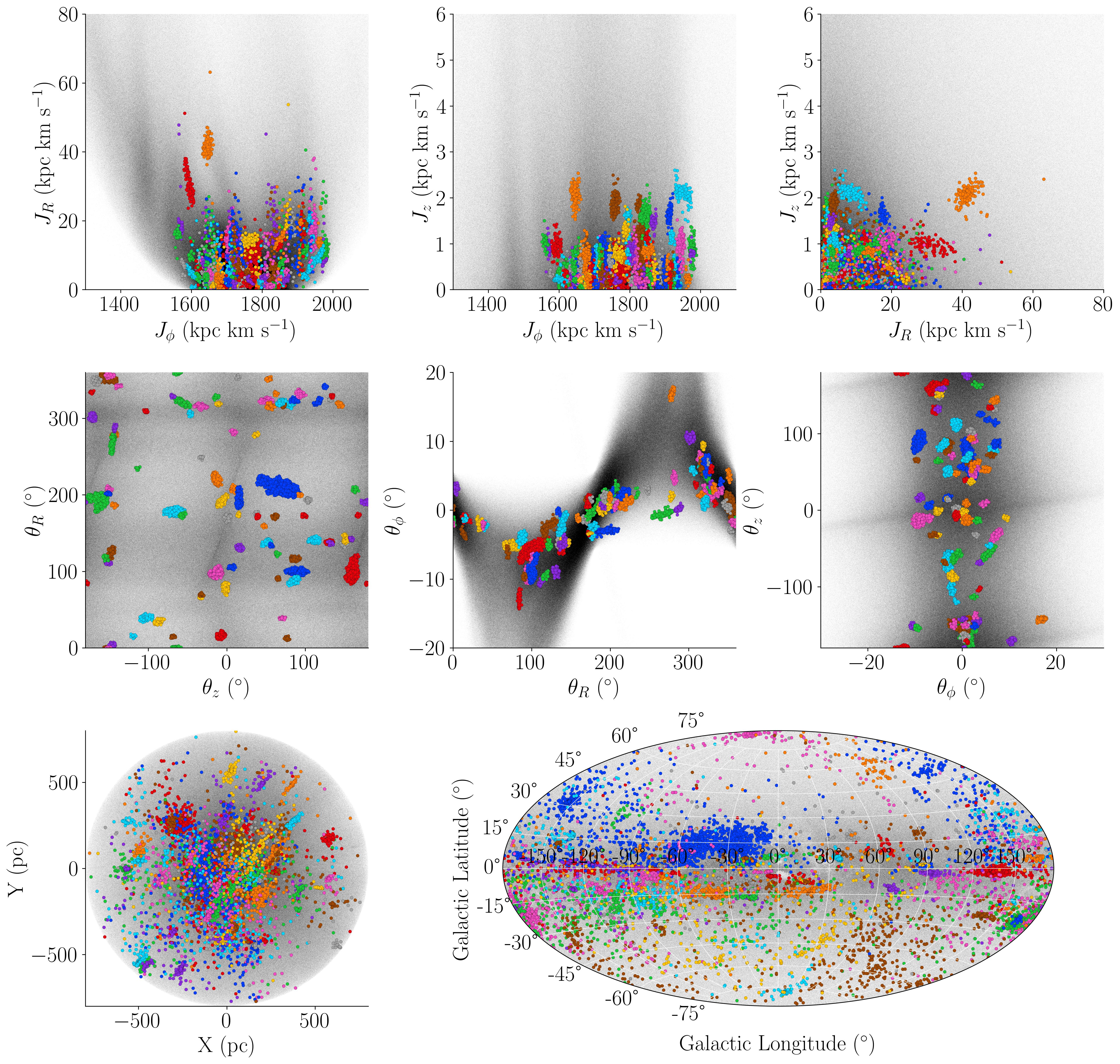}
\caption{The member stars of the 94 most prominent action-angle groups identified with DBSCAN. The various colors represent the different groups. The black background corresponds to all sources in the \textit{Gaia RVS sample}. Top/Middle row: The distribution of the groups in the three projections of the action/angle space. By construction, all groups are tightly clustered in this parameter space. Bottom left: Top-down view on the 94 groups in Galactic Cartesian coordinates, where the Sun is located at (X,Y) = (0,0)~pc. Bottom right: The sky distribution of the action-angle groups in Galactic longitude and latitude, illustrated in a Hammer projection. Although all groups are very tightly clustered in action-angle space, most of them are extended across several hundred degrees on the sky. \label{fig:groups}}
\end{figure*}

\subsection{Identifying Stellar Overdensities in 6D Action-Angle Space} \label{clustering}

\begin{figure*}[ht!]
\includegraphics[width=\textwidth]{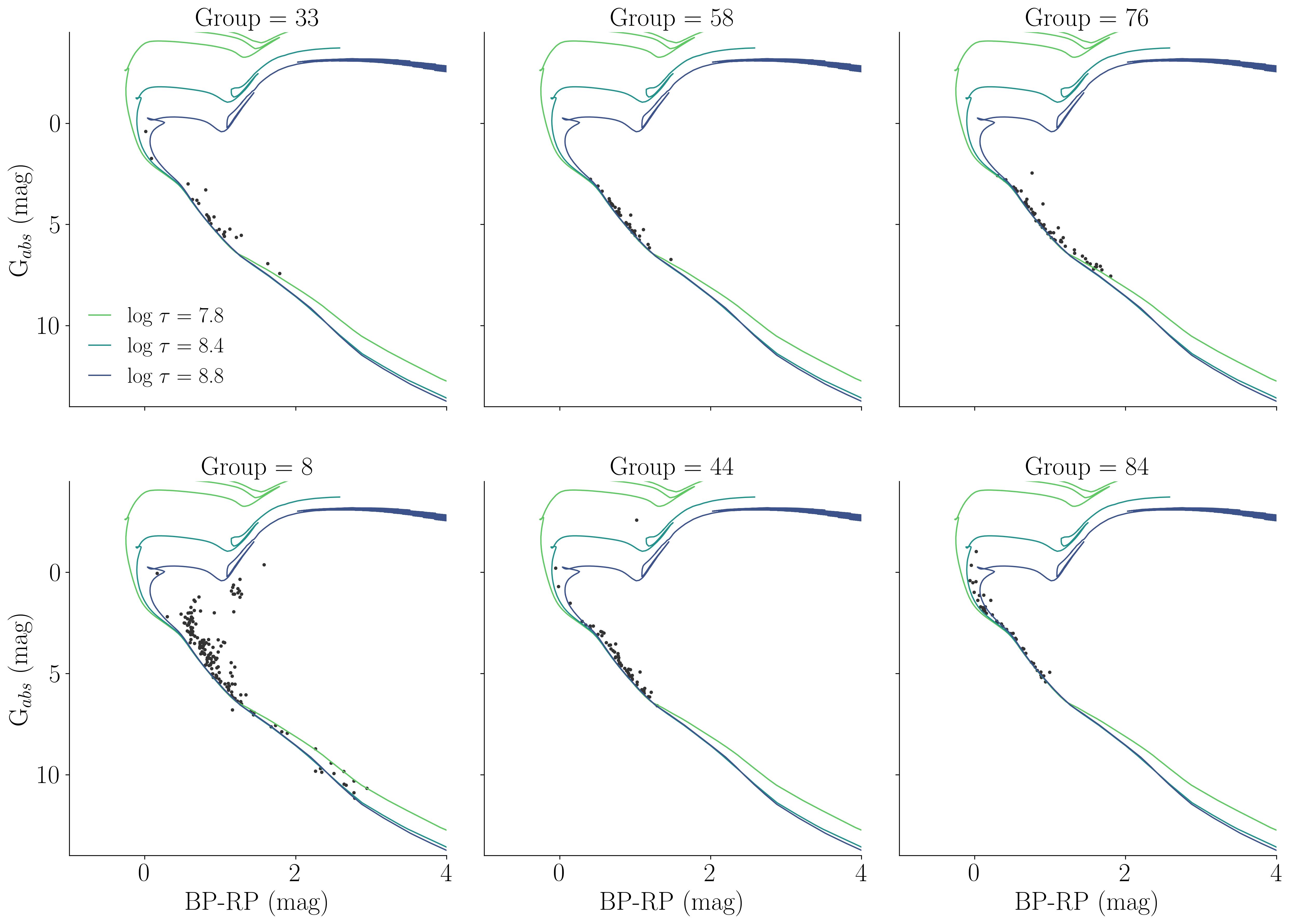}
\caption{Observational CMDs for six representative action-angle groups. The member stars of each group are arranged in narrow, but short, sequences, where stars are clearly missing in the upper and lower main-sequence parts. This is caused by the limited magnitude and temperature range of the \textit{Gaia RVS sample}, where only stars with G $<$ 14~mag and with effective temperatures in the range of 3.100 to 14.500~K are included. In color, we show a set of three isochrones (log $\tau$ = 7.8, 8.4, 8.8), illustrating that stars located on the main sequence only have a small age-diagnosing value. \label{fig:CMDs_short}}
\end{figure*}

In \citet{coronado2022} we showed that stellar overdensities -- from clusters, to associations, to disk streams -- can be identified as groups that are overdense in action-angle space. Here, we follow the same approach applied to our \textit{Gaia RVS sample}, with modifications that we summarize here. First, we describe all orbital angles via their \emph{sine} and \emph{cosine} functions bypass complexities that arise from the periodicity of these variables. To reach a consistent metric across all six coordinates that serves our purpose, we first normalize each of the action-angle coordinates (J$_{R}$, J$_{\phi}$, J$_{Z}$, sin($\Theta _{R,\phi,Z}$), cos($\Theta _{R,\phi,Z}$)) by removing the mean and scaling to unit variance. Additionally, we explore changes in the relative weight of "orbits" vs "orbital angles", and reduce the variance in the six angle coordinates (now sines and cosines of the angles) by multiplying the scaled data with 0.4.

To identify clusters of objects in this 6D space, \citet{coronado2022} had used a version of the Friends-of-Friends algorithm to link objects. Here, we apply the density-based algorithm DBSCAN \citep{ester_1996} to extract clustered ensembles of objects from the (J$_{R}$, J$_{\phi}$, J$_{Z}$, sin($\Theta _{R,\phi,Z}$), cos($\Theta _{R,\phi,Z}$)). We choose this clustering algorithm instead of the Friends-of-Friends algorithm because it is much faster in handling large datasets. Running DBSCAN requires two choices: a minimum number of members (minPts), and a maximum distance scale $\epsilon$. We have explored a wide range of $\epsilon$ and minPts parameters, and decided on a final setup of $\epsilon$ = 0.035 and minPts = 20. In combination with our orbital angle weighting, this parameter choice yields to 0.1\% of the stars in the \textit{Gaia RVS sample} to be clustered in about twice as many groups as we had identified in \cite{coronado2022}. With these choices we now identify 102 action-angle groups, and keep the 94 most prominent groups that have at least 20 member stars. This sample is large enough to develop, test, and apply our age estimates; and it is large enough to get a first census which fraction of groups are ``young'' and approximately ``mono-age``. 

Fig.~\ref{fig:groups} illustrates these 94 action-angle groups: the top row shows their distribution (color-coded by group) in the three action plane projections, and the middle row shows their distribution in the three angle plane projections. By construction, all groups are very compact in this parameter space. The bottom row shows a top-down view on the 94 groups in Galactic Cartesian Coordinates (left), and their distribution on the sky in Galactic longitude and latitude (right). Although the groups are tightly clustered in action-angle space, most of them extend many degrees on the sky, and a substantive fraction over several 100 parsecs in physical extent.

\begin{figure*}[ht!]
\includegraphics[width=\textwidth]{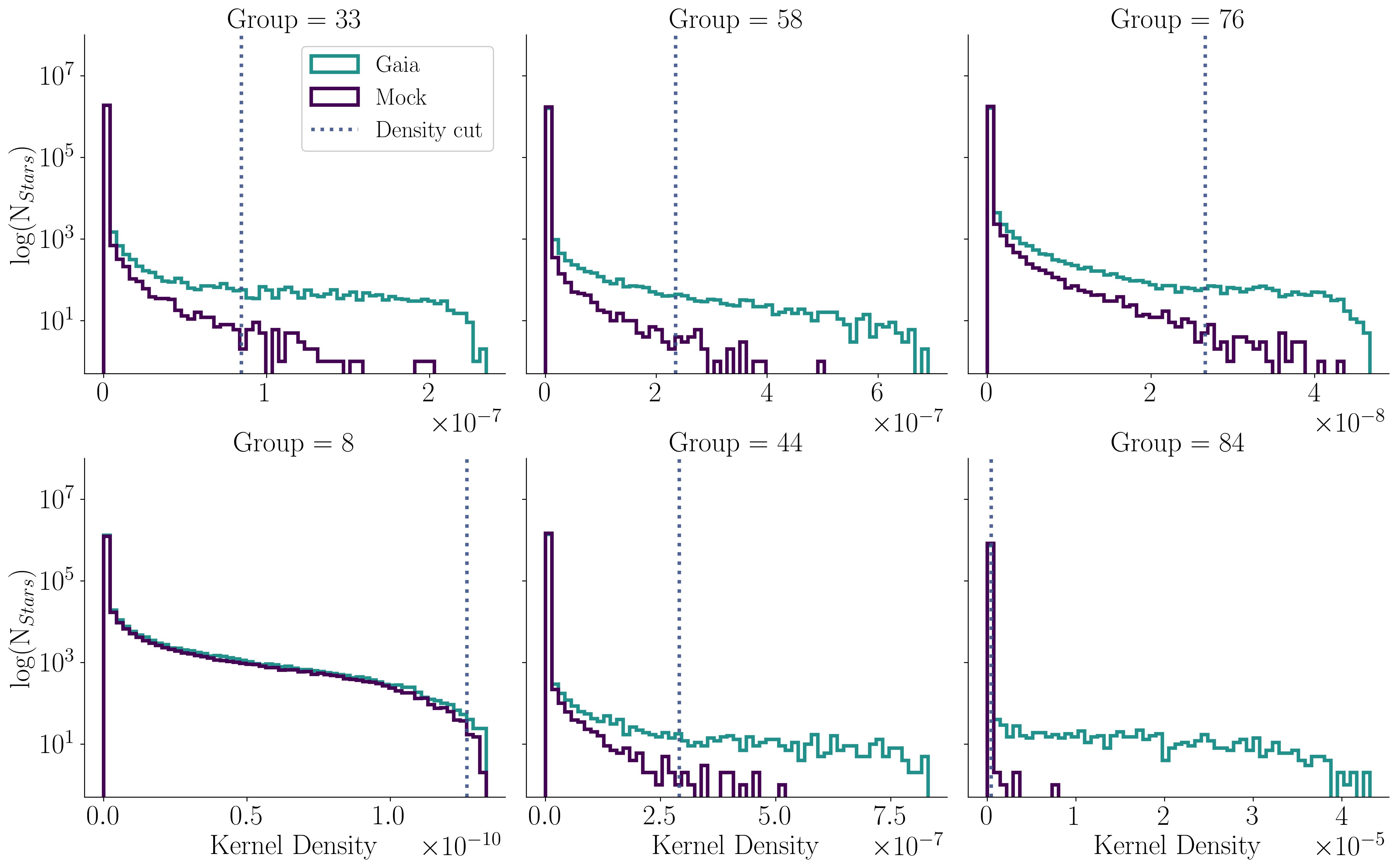}
\caption{Histograms of the allocated density values for all Gaia (turquoise) and Gaia mock stars (violet) within the individual (X, Y, Z, $\mu_{\alpha}$, $\mu_{\delta}$) ranges} for six representative action-angle groups. The dotted blue line marks the location of the final density cut T$_{final}$ as described in Subsection~\ref{5D}. Most Gaia and Gaia mock stars have a density value close to zero, with a tail extending to higher densities, where typically the amount of Gaia stars exceeds the amount of Gaia mock stars. This is an indication that we find stars clumped in position-proper motion space in the \textit{Gaia sample}.
\label{fig:histograms}
\end{figure*}

For six representative action-angle groups Fig.~\ref{fig:CMDs_short} displays the CMDs for their members identified in 6D space. In all cases, these members form a narrow sequence in CMD space, as expected for simple stellar populations. However, the overlaid isochrones in Fig.~\ref{fig:CMDs_short} also illustrate that these stars only cover a portion of the expected isochrone locus.  This is for two reasons: the limited magnitude range and the limited temperature range of the \textit{Gaia RVS sample}, where only stars with G $<$~ 14 mag and effective temperatures in the range of 3.100 to 14.500~K are included. For the age determinations of these groups this is a serious limitation, as the (upper) main-sequence turn-off (for ages $>10^8$~ yrs) and the lower main sequence (for ages $<10^8$~yrs) provide the best age constraints for stellar ensembles. 

This is apparent from the set of three isochrones overplotted in Fig.~\ref{fig:CMDs_short}, spanning an age range of $\tau$ = 60 - 630~Myrs, specifically log $\tau$ = 7.8, 8.4, 8.8. E.g. for Group 58 in the top row, the group members identified in 6D are located exclusively on the part of the main sequence where all these isochrones coincide.

Hence, determining ages by isochrone fitting requires identifying member stars at temperatures (or stellar masses) that are not included in the \textit{Gaia RVS sample}. Such hotter and cooler stars that are missing here should, however, be in the overall \textit{Gaia sample}, with valid 5D position -- proper motion entries. 

\subsection{Associating additional member stars in the 5D space of positions and proper motions} \label{5D}

In this Section we describe how to extend the stellar membership of the action-angle groups across a wider color or stellar-mass range, in order to overcome the just-described limitations in age dating these groups. We do this by mapping their distribution from 6D (rescaled) action-angle space to 5D position-proper motion space (X, Y, Z, $\mu_{\alpha}$, $\mu_{\delta}$), to search for the most likely members in this projected space, where we do not have the color limitations of the \textit{Gaia RVS sample}.

 We start by constructing an approximate Probability Density Function (PDF) for each group in 6D space. We do this with the help of a kernel density estimate for each group in the rescaled 6D space of J$_{R}$, J$_{\phi}$, J$_{Z}$, sin($\Theta _{R,\phi,Z}$), cos($\Theta _{R,\phi,Z}$), using a Gaussian kernel with a unit bandwidth that we can then sample finely. Before doing that, we remove members in each action-angle group that are beyond 2$\sigma$ from the median in any coordinate, to mitigate the impact of outliers; this leaves about 95\% of the stars.
 We then sample this PDF estimate with 5000 random points that we can then map from action-angle space (J$_{R}$, J$_{\phi}$, J$_{Z}$, $\Theta_{R}$, $\Theta_{\phi}$, $\Theta_{Z}$) to position-proper motion space ($\alpha, \delta, \varpi, \mu_{\alpha}, \mu_{\delta}, v_{r}$). For this mapping we use the MWPotential2014 in the TORUSMAPPER package \citep{binney_2016}, which is part of the AGAMA library \citep{vasiliev_2019}, a software framework offering several tools applicable to galaxy modeling and stellar dynamics. We then remove some "nonphysical" sampled datapoints, e.g. datapoints with negative radial and vertical actions.
 
Finally, we take the sampled datapoints that have been mapped into 5D space and perform a kernel density estimate, where we choose  (X, Y, Z, $\mu_{\alpha}$, $\mu_{\delta}$) as coordinates, to have an intuitive metric for the spatial part. In each dimension, we use a Gaussian kernel with a unit bandwidth to obtain a density estimate. This smooth 5D density now allows us to assign a local density value $\rho$ to each star in the actual \textit{Gaia sample} that is within the groups individual boundaries{, e.g. minimum and maximum coordinates in (X, Y, Z, $\mu_{\alpha}$, $\mu_{\delta}$).}

Of course the 5D space just described can and will also contain unrelated field stars. We can only identify new members of this group with any confidence, where the 5D density of the actual \textit{Gaia sample} is higher than the expectation from a smooth mock catalog. Therefore we apply the same procedure to the smooth \textit{Gaia mock sample} described in Section~\ref{mock}, leaving us with a local density value assigned to each star in the mock sample.

Now we can identify the domain of this 5D space where each group's 5D density is far higher than the density in the smooth \textit{Gaia mock sample}. To this end, we create a histogram of the number of Gaia ($n_{gaia}$) and Gaia mock ($n_{mock}$) stars as a function of their assigned Kernel density $\rho$. We set a density threshold for each group with the following approach: The first density threshold T$_{1}$ is the lowest density bin where the local density of the \textit{Gaia sample} is 25 times higher than expected from the \textit{Gaia mock sample}. Given that the \textit{Gaia mock sample} is an imperfect representation of the real data, we decide on the ratio number 25 after some experimentation. This selects all stars in a group as likely 5D members, if their local density is 25 times higher than expected from the \textit{Gaia mock sample}. Second, we calculate the ratio between $n_{gaia}$ and $n_{mock}$ in each bin and set the second density threshold T$_{2}$ at the density with the highest contrast, with the additional condition to contain at least 100 stars. We pick a final density threshold T$_{final} = min(T_1, T_2)$ for each group by picking the threshold at the lower density (minimum), e.g. to assign as many 5D candidate members as possible to each group.

We find that about half of the groups (41) get the density threshold T$_1$ assigned, whereas the remaining groups (53) get the density threshold T$_2$ assigned. From the groups with T$_2$, 16 groups get the density threshold at 100 stars, and 37 groups get a density threshold at lower densities with a higher ratio.

For each group we then merge our original group member selection drawn from the \textit{Gaia RVS sample} with the new group member selection obtained from the full \textit{Gaia sample}. From this final group member selection, we manually remove two groups that show overpopulated CMD's and end up with a final set of 92 groups with extended memberships to test and apply the age dating method described in Section~\ref{sec:ages}. The full "group member selection" is available at the CDS, including the Gaia DR3 Source ID, the Group membership, and the coordinates $\alpha, \delta$.

\begin{figure*}[ht!]
\includegraphics[width=\textwidth]{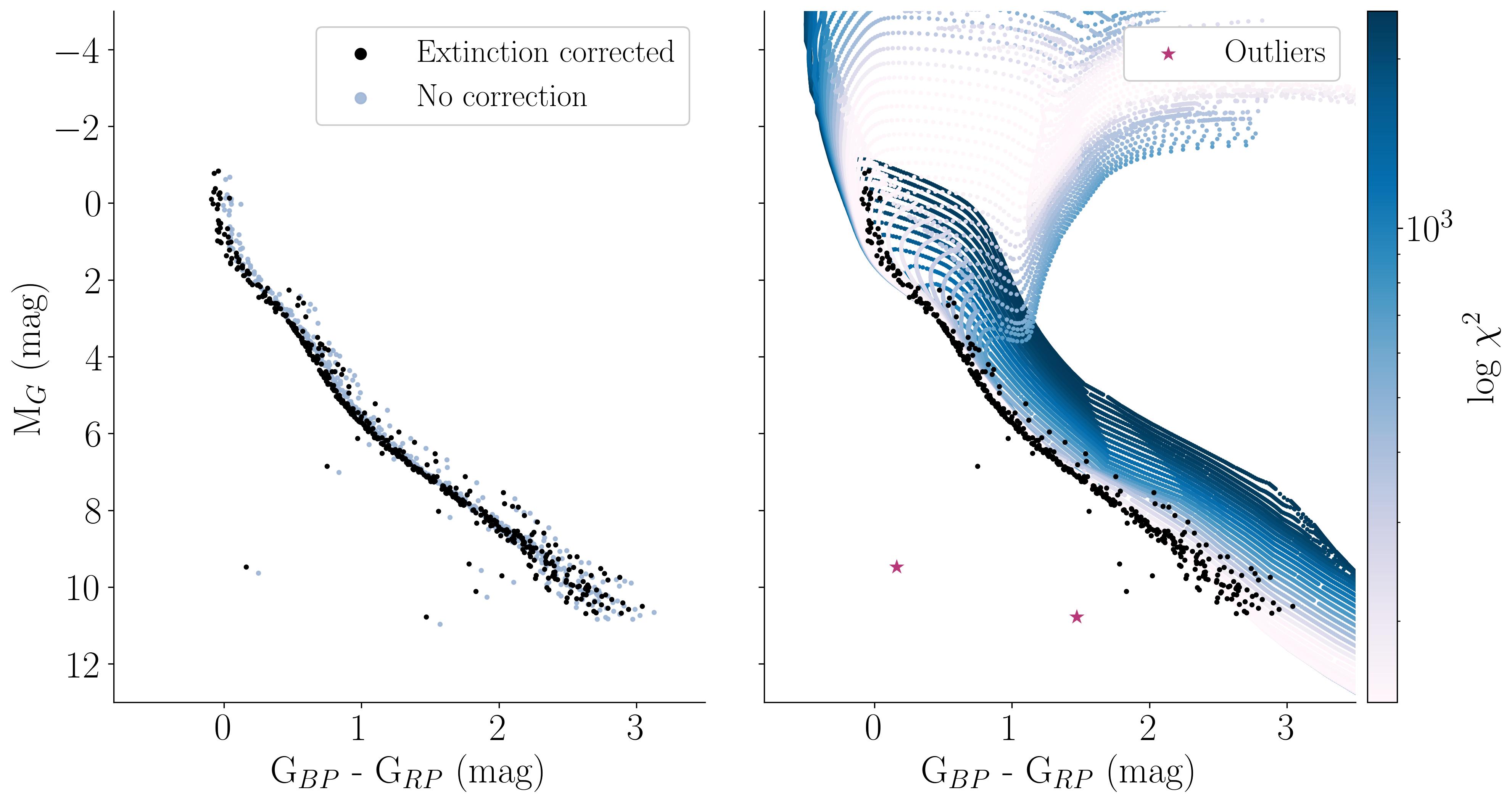}
\caption{Left: The effect of the extinction correction method for stars in Group 68. The correction reveals a significant shift of the uncorrected (colored) stars to the reddening corrected (black) stars, emphazising the importance of extinction correction for precice age dating via isochrone fitting of CMD's, even for stars located in the solar neighborhood. The average shift in color for Group 68 is 0.1~mag. Right: Illustration of the likelihood isochrone fitting for Group 68. All isochrones are colored by age (107 different logarithmic $\tau$) for a fixed metallicity of [Fe/H] = 0.1. The color coding represents the $\chi^2$ value obtained for the different age setups. The plot illustrates that very young as well as very old ages are not likely to fit the group's distribution, as they lead to a high $\chi^2$ value. The best fit isochrone for this group is the one with the lowest $\chi^2$ value. The stars in magenta mark group members that are labelled as 4-$\sigma$ outliers by our code setup, and are not considered in the likelihood isochrone fitting. We show the isochrones not as solid lines, but as datapoints to illustrate our finely spaced magnitude and color setup, which we obtain by linear interpolation (see Sec.~\ref{subsec:iso}).} \label{fig:method_2}
\end{figure*}

The above described procedure is illustrated in Fig.~\ref{fig:histograms}. It shows the density histograms for all Gaia stars (turquoise) and Gaia mock stars (violet)  for six representative action-angle groups, selected within the individual group's volume and proper motion range. Most Gaia and Gaia mock stars have small density values, with a tail extending to higher densities. As expected, the actual Gaia stars -- in this 6D group selected patch -- far exceed the number of Gaia mock stars in the high density regime. This illustrates immediately that we find these groups also as prominent overdensities in the 5D \textit{Gaia sample}. The dotted blue line marks the location of the density cut T$_{final}$, which varies considerably between the groups, depending on the distribution of the Gaia mock stars. Some groups exhibit a more pronounced excess of Gaia stars in the high density regime than other groups. This could be for two reasons: either some groups might be more compact, or some groups may occur far away from the disk plane, which both provides naturally a higher density contrast.

\section{Age determination} \label{sec:ages}

We now want to determine stellar ages of the 92 action-angle groups. In the end this estimate comes from a forward model of the photometry, $\vec{data}$, of the group members, $p\bigl( \tau | \{\vec{data}\}\bigr )$, where $\vec{data}$ incorporates the magnitudes and parallaxes of the stars, as well as their errors. The magnitude errors are not provided in the Gaia catalog, therefore we obtain them from the G, BP and RB band fluxes and flux errors. In this modelling, we want to account for the following effects: The stars are widely spread across the sky (see Fig.\ref{fig:groups}), they may have widely ranging member distances (we obtain distances by inverting the parallax), and they have varying photometric and distance uncertainties. In addition, we need to allow for (unresolved) binaries, possible outliers, and we need to take into account differential reddening.  

\begin{figure*}[ht!]
\includegraphics[width=\textwidth]{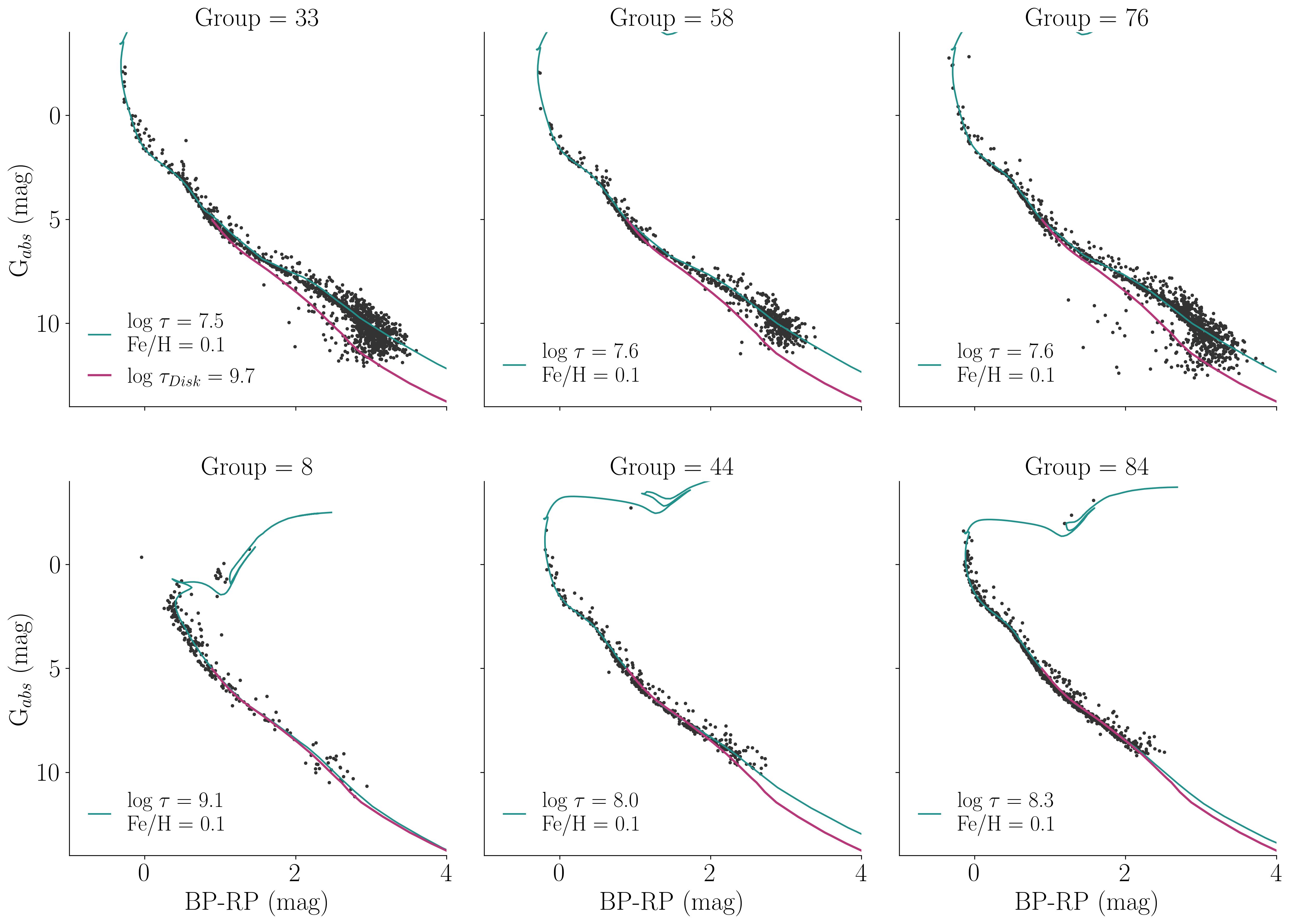}
\caption{Observational CMD's for six representative groups, with additional member stars associated from the 5D position-proper motion space. All stars are individually extinction corrected. The turquoise curve on top of each sequence shows the best-fitting isochrone (log $\tau$, [Fe/H]). Across all groups, the isochrones fit both the upper and lower main sequences very well. The magenta isochrone at the lower main-sequence represents the age of the Galactic disk. The three groups in the top row are very young ($<$ 25~Myrs), and their lower main sequences are clearly offset from the age of the Galactic disk. This again illustrates the power of the lower main sequence to determine the ages of very young populations. The three groups in the bottom row are older, and for them the main-sequence turn-off is the main age determinant. \label{fig:CMDs_long}}
\end{figure*}

\subsection{Reddening correction} \label{reddening}

We start with the reddening correction for each individual star. It is tempting to simply take each group member's reddening estimate directly from the \textit{Gaia extinction sample}. These estimates are based on the idea to deredden each star until it gets to its most plausible place in the CMD. However, for stars that lie above the solar-metallicity, zero-age main sequence -- whether they are young stars or binaries -- this often leads to overestimates of their reddening. Instead, we adopt the ensemble median of other stars in the same direction and at the same distance. To obtain robust estimates, we limit these reddening estimates to stars in the \textit{Gaia extinction sample} where we can assume that they are firmly on the zero-age main sequence, requiring 2~mag~$<$~M$_G$~$<$ 8~mag. We select all stars that are located within a sphere of radius 25~pc around each group member star under consideration for dereddening, and pick the 50$^{th}$ percentiles in A$_{G}$, A$_{BP}$, and A$_{RP}$ to subtract from the member star's uncorrected Gaia magnitudes G, BP, and RP. Depending on the location of each group and the location of each stars within the group, the spheres contain between 266 and 2059~stars, typically around 800~stars. The effect of this extinction correction is illustrated for Group 68 in the left panel of Fig.~\ref{fig:method_2}. The colored data points represent the CMD before applying the reddening correction, and the black data points show the reddening corrected sequence. E.g. for Group 68, the average shift in color is 0.1~mag. Fig.~\ref{fig:method_2} illustrates that reddening correction is important for precise isochrone fitting, even for stars located within the extended solar neighborhood of 800~pc from the Sun.

\begin{figure}[t!]
\includegraphics[width=\columnwidth]{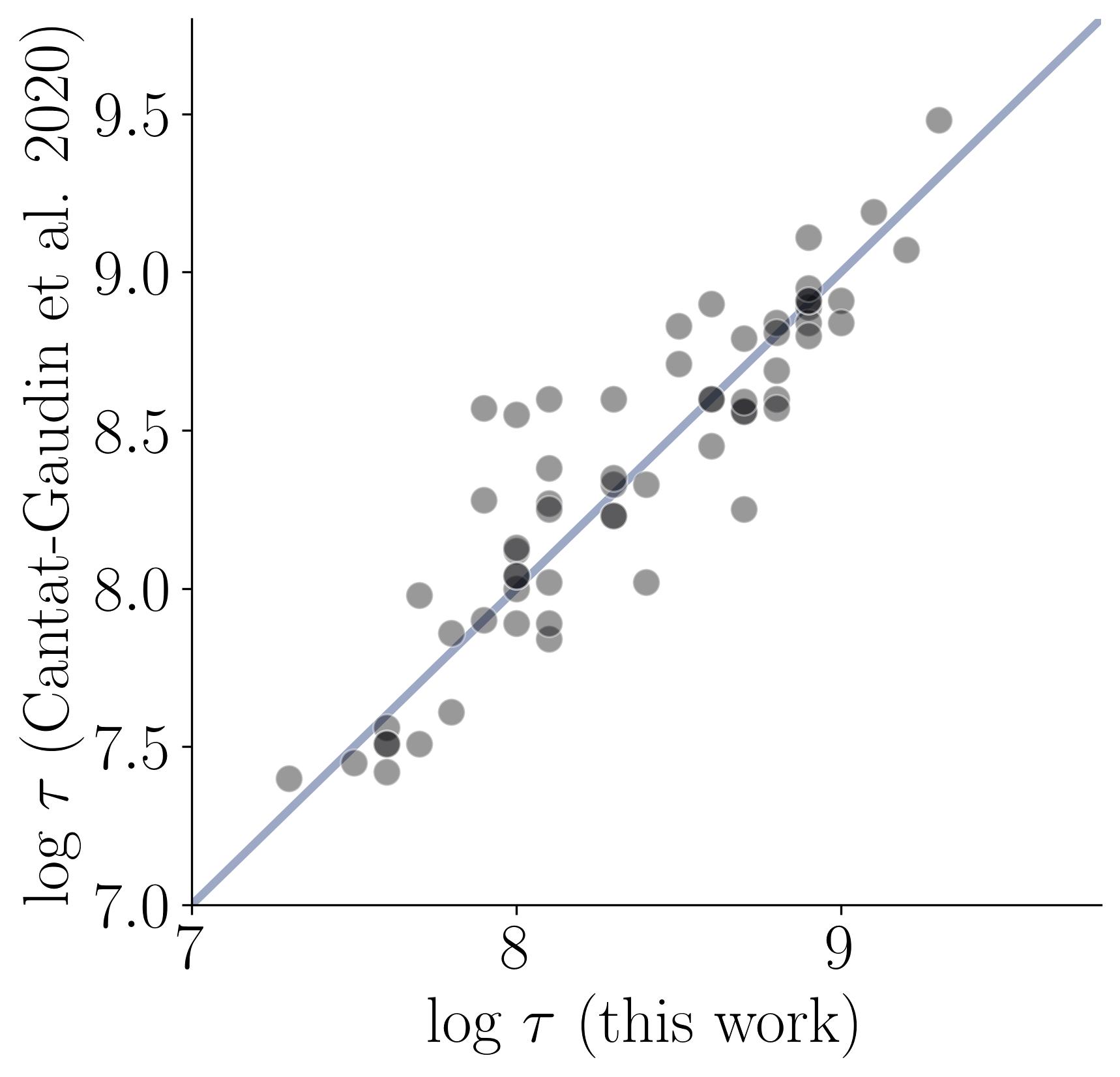}
\caption{Comparison between the age determination of this work with the age determination of \citet{cantatgaudin_2020} for 62 established clusters that we could identify in the cluster catalog by \citet{cantatgaudin_2020}. The plot shows that the age estimates from this work are in very good agreement with the literature values, with only very little scatter
The $\ltau$ uncertainties of the age fit range from 0.1 - 0.25 for \citet{cantatgaudin_2020}, and are typically smaller than 0.1 for the method presented in this work. \label{fig:literature}}
\end{figure}

\subsection{Basic Fitting Methodology} \label{likelihood}

We now perform isochrone fitting by calculating the likelihood (here $\chi^2$) of the reddening-corrected photometric data and parallax for each group, 
$\{\vec{data}\}_{i=1,N_{\star}}$. Here, $N_{\star}$ is the number of likely 5D members in the group and each star's data is given by its apparent magnitudes and parallaxes, as well as their errors: $\vec{data} \equiv \bigl (m_{G},m_{BP},m_{RP}, \varpi;~ \sigma_{m_{G}},\sigma_{m_{BP}},\sigma_{m_{RP}},\sigma_{\varpi}\bigr )$. 
Each isochrone, specified by the two parameters log $\tau$, [Fe/H], provides model predictions $M_{band} = f\bigl (\mathcal{M}~|~\mathrm{log}\tau,\mathrm{[Fe/H]}\bigr )$, where \textit{band} stands for G, BP and RP, and $\mathcal{M}$ is the initial stellar mass. 

A priori, we neither know any star's mass nor its exact distance, and we should treat these quantities as parameters, even if for the current context they are ``nuisance parameters''. This means we have to  optimize \emph{both} the global parameters $\mathrm{log}\tau,\mathrm{[Fe/H]}$ and the individual masses $\mathcal{M}$ and distance moduli (DM), which are mainly constrained by the parallaxes, $\varpi$.  We perform this optimization by a multi-dimensional grid search. To speed up the computation we only consider a very limited range of DM--$\mathcal{M}$ combinations for each star, constrained by their apparent magnitude and parallax. 

We only consider distance moduli that are $2\sigma$ consistent with the parallax measurement:

\begin{equation}
\label{eqn:distancerange}
    \begin{aligned}
        DM_{min} = 10 - 5 \cdot \log_{10}(\varpi + 2 \cdot \sigma_{\varpi}) \\
        DM_{max} = 10 - 5 \cdot \log_{10}(\varpi - 2 \cdot \sigma_{\varpi}), \\
    \end{aligned}
\end{equation}

and we only consider (pre-computed) $M_G(\mathcal{M})$ in the range 

\begin{equation}
\label{eqn:massrange}
    \begin{aligned}
        M_{G_{min}} = (m_{G} - 2 \cdot \sigma_{m_{G}}) + 5 \cdot \log_{10}(\varpi - 2 \cdot \sigma_{\varpi}) \\
       M_{G_{max}} = (m_{G} + 2 \cdot \sigma_{m_{G}}) + 5 \cdot \log_{10}(\varpi + 2 \cdot \sigma_{\varpi}).
    \end{aligned}
\end{equation}

Finally, to treat the parallax measurements as data we need to have a parallax prediction, which is simply

\begin{equation}
\label{eqn:predictedplx}
    \begin{aligned}
        \varpi_{pred} = 10^{-\frac{(DM - 10)}{5}}.
    \end{aligned}
\end{equation}

On this bases, we can calculate for each assumed parameter set and each star $\chi^{2}\bigl (\mathrm{DM},\mathcal{M} ; \mathrm{log}\tau,\mathrm{[Fe/H]}\bigr )$:

\begin{equation}
\label{eqn:chi2}
    \begin{aligned}
        \chi^{2} =  \frac{(m_{G} - DM - M_{G})^{2}}{\sigma_{m_{G}}^{2}}+\frac{(m_{BP} - DM - M_{BP})^{2}}{\sigma_{m_{BP}}^{2}}+ \\
        \frac{(m_{RP} - DM - M_{RP})^{2}}{\sigma_{m_{RP}}^{2}}+\frac{(\varpi - \varpi_{pred})^{2}}{\sigma_{\varpi}^{2}}
    \end{aligned}
\end{equation}

For any set of $\mathrm{log}\tau,\mathrm{[Fe/H]}$ (the outer loop in the optimization) we optimize $\mathrm{DM},\mathcal{M}$ for each star, and then sum all stars' $\chi2$ to obtain $\chi^2_{tot}\bigl ( \mathrm{log}\tau,\mathrm{[Fe/H]} \bigr )$. This procedure then results in a mapping of the whole data set for any given group as a function of $\mathrm{log}\tau,\mathrm{[Fe/H]}$, including their best fit values. 

The right panel in Fig.~\ref{fig:method_2} illustrates the outcome of this process for Group 68. In color are shown isochrones for 107 different log $\tau$, at a fixed metallicity of [Fe/H] = 0.1. The color coding represents the $\chi^2$ value obtained for the different age setups. The plot illustrates that the ages are well constrained both the upper and lower main sequence, as both regimes lead to a high $\chi^2$ value. 

This procedure is straightforward and accounts for the varying distances, reddenings and data uncertainties (including the parallax). We now turn to the generalization of this procedure to incorporate outliers and to consider the possibility of an intrinsic age spread in the stellar groups.

\begin{figure}[ht!]
\includegraphics[width=\columnwidth]{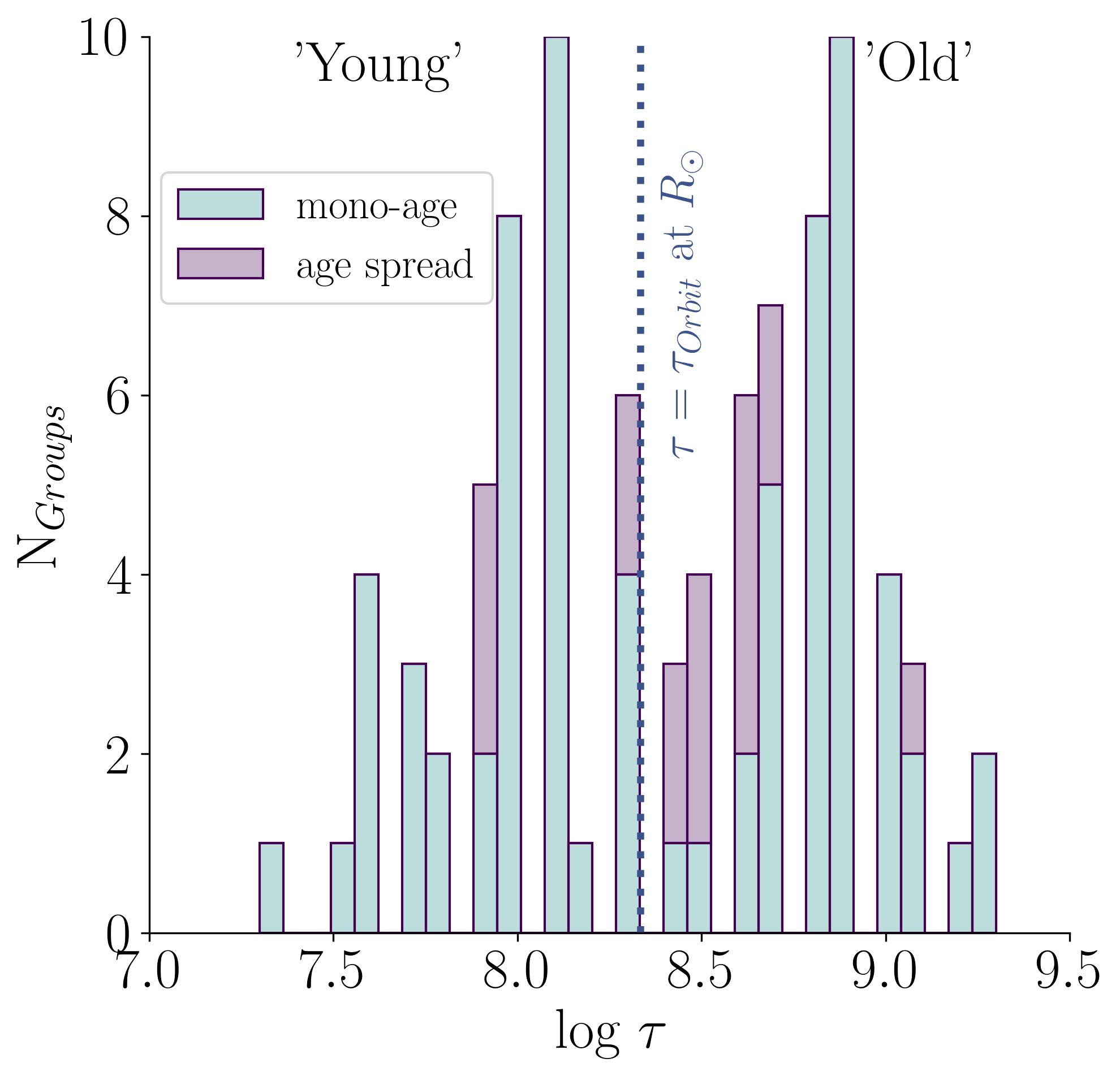}
\caption{Histograms of the distribution of age (log $\tau$) across the set of 92 groups. About half of them (43) are younger than an orbital period, and 49 of them are older. The central timescale, the orbital period at the Solar radius is indicated by the blue dotted vertical line, for which we adopt $\mathrm{log}\tau = 8.33$. The intrinsic age spread is indicated in the colorcoding of the bars, where groups with an intrinsic age spread are shown in magenta, and mono-age groups are illustrated in turquoise.}
\label{fig:values}
\end{figure}

\subsection{Accounting for Binaries and Outliers} \label{outbin}

Isochrones reflect the color-magnitude locus of single stars, while observed CMDs have ``outliers'' for a number of reasons. Such outliers may be spurious data or stellar systems whose physics is not captured by the isochrone: (unresolved) binary stars, evolved stars such as white dwarfs, or actual interlopers of different ages and metallicities. In addition, the color prediction of the lowest mass stars may be less robust for isochrone models, which may explain some of the deviations.
 
In our code, we implement two different approaches to deal with binaries on the one hand, and different outliers on the other. 
For binaries, we follow \citet{coronado_2018}, where we generalize the isochrone prediction of a star's absolute magnitude $M_{iso}$ to be either that of a single star within a magnitude width $\sigma_M$, or -- with probability $\epsilon$ -- to be that of a near-equal mass binary that is 0.7 mag brighter, or to be a binary of any kind between $M_{iso}$ and $M_{iso}$ + 0.7 mag. Note that some groups illustrated in Fig.~\ref{fig:CMDs_long} exhibit well populated binary sequences, illustrating the need for this methodological extension. 

This results in a probability 

\begin{equation}
    \label{eq:binary}
   \begin{aligned}
    p\bigl( M_{\lambda} | \mathcal{M} ; \mathrm{log}\tau,\mathrm{[Fe/H]}, \epsilon\bigr )=\\ (1-\epsilon)~\mathcal{G}\bigl( M_{\lambda}- M_{\lambda,iso},\sigma_{M} \bigr) + \\
    \frac{\epsilon}{2}~\Bigl( \mathcal{G}\bigl( M_{\lambda} - (M_{\lambda,iso}-0.7),\sigma_{M} \bigr) + I(M_{\lambda})\Bigr ).
   \end{aligned}
\end{equation}

where $\mathcal{G}$ is a Gaussian and $I$ is the ``in-between'' term defined as $1/0.7$ between $M_{iso}$ and $M_{iso}+0.7$.

The logarithm of p(M) in Eq.~\ref{eq:binary} then simply gets added for each object to the $\chi^{2}$ in Eq.~\ref{eqn:chi2}.

We deal with the other outliers by limiting all of them to be ``n-$\sigma$ outliers'' at worst: any individual $\chi^2$-term cannot be larger than $2\times n^2$. We adopt $n=4$, or $\chi^2_{max}=32$.  This is illustrated in Fig.~\ref{fig:method_2} for example Group 68, where stars in magenta mark group members that are labeled as 4-$\sigma$ outliers by our code setup, and are not considered in the likelihood isochrone fitting.

\subsection{Incorporating Age dispersion} \label{dispersion}

The approach described so far determines ages, assuming a priori that we have mono-age populations. However, we also want to consider the case that groups may have some finite age dispersion:
$p\bigl (\mathrm{log}\tau~|~\overline{\mathrm{log} \tau},\:\sigma_{\mathrm{log}\tau}\bigr ) \equiv \mathcal{G}\bigl(\mathrm{log}\tau-\overline{\mathrm{log}\tau},\:\sigma_{\mathrm{log}\tau} \bigr )$, where the mean (log) age is $\overline{\mathrm{log}\tau}$ and its dispersion is $\sigma_{\mathrm{log}\tau}$. We implement this by an extension of Eq.~\ref{eqn:chi2}: We consider the ``linearized'' version of $\chi^2\bigl ( \mathrm{log}\tau,\mathrm{[Fe/H]} \bigr )$, i.e. $\exp{\chi^2( \mathrm{log}\tau,\mathrm{[Fe/H]})}$, and then integrate over all $\mathrm{log}\tau $ within $\pm 3 \sigma_{\mathrm{log}\tau} $:
\begin{equation}
e^{\chi^2_{tot}}\equiv \int e^{\chi^2\bigl ( \mathrm{log}\tau|\mathrm{[Fe/H]} \bigr )} \, p\bigl (\mathrm{log}\tau~|~\overline{\mathrm{log} \tau},\:\sigma_{\mathrm{log}\tau}\bigr ) \mathrm{dlog}\tau .
\end{equation}

In this case we optimize globally over three parameters, the mean age log$\tau$, the metallicity [Fe/H], and the intrinsic age spread $\sigma_{\mathrm{log}\tau}$.

We then estimate the uncertainty $\delta_{\ltau}$ of the isochrone fit for each group and compare it to the individual intrinsic age dispersions $\sigma_{\ltau}$, in order to determine what value of $\sigma_{\ltau}$ is significant ("effectively mono-age"). We use the $\chi^2$ values to determine the data probabilities and then look at the marginalised distributions to get the uncertainties in the individual parameters. Rather than quoting them individually, we find that the fit uncertainty $\delta_{\ltau}$ is in 90$\%$ (83) of the cases smaller than 0.1, which is less than our $\ltau$ grid size of 0.1. In 9 cases the fit uncertainties are larger than 0.1, whereas in only 1 case $\delta_{\ltau}$ is larger than 0.15. Therefore, we choose a value of $\sigma_{\ltau}$ = 0.2 to declare "effectively mono-age" as significant.

\section{Results} \label{sec:results}

With these isochrone fitting tools in hand, we can now proceed to presenting the results from the CMD's for 92 groups described in Sect.~\ref{sec:groups}.

\subsection{Illustrating the Age Constraints from Isochrone Fits} \label{iso_illustration}

We start by illustrating the results derived from the above procedure for the same six groups from Fig.~\ref{fig:CMDs_short}, now shown in Fig.~\ref{fig:CMDs_long}, with the full set of 5D proper motion members: the isochrones are now populated over a wider age range, providing better age constraints. All stars have been individually reddening corrected. 

On top of each group the best fitting single-age isochrones are shown in turquoise (log $\tau$, [Fe/H]), derived following Section~\ref{sec:ages}. Across all groups, the isochrones fit both the upper and lower main sequences very well. In particular, this Figure illustrates that neither the binary sequence nor any outliers (e.g. below the main sequence) have noticeably biased the fit. The magenta isochrone at the lower main-sequence represents the mean age of the Galactic disk (log $\tau_{Disk}$ = 9.7). 

The three groups in the top row are very young ($<$ 25~Myrs), and low-mass stars have clearly not yet reached the true zero age main sequence. This illustrates well the power of low-mass stars for constraining very young population ages, for which the main sequence turn-off is sparsely sampled. The three groups in the bottom row are older. For them low-mass stars provide no two-sided age constraint. But for them the main-sequence turn-off is at lower masses, hence well-population and providing the main age constraint. 

While these fits are pleasing, they need to be validated against external results, which we do by comparing them to the ages from \citet{cantatgaudin_2020}. We crossmatch our groups with known clusters in the \citet{cantatgaudin_2020} catalog. We keep all groups for which we obtain at least 20 crossmatches with one literature cluster. Additionally, for groups that have matching sources with more than one literature cluster, we keep the literature cluster with the highest number of -- but at least 20 -- crossmatches. This applies to 62 of our groups. Figure~\ref{fig:literature} compares our age determinations to those of \citet{cantatgaudin_2020}, showing overall very good agreement, with no systematic bias and very little scatter. The $\ltau$ 
 uncertainties for individual clusters in the \citet{cantatgaudin_2020} catalog range from 0.1 - 0.25, depending on the age itself and the number of stars \citep{cantatgaudin_2020}. As described in more detail in Section~\ref{dispersion}, the $\ltau$ uncertainties of our isochrone fitting approach are typically smaller than 0.1, which is less than our $\ltau$ grid of 0.1.

\subsection{Ages, Age Dispersions, [M/H] and Spatial Structure of the Action-Angle Groups} \label{properties}

These fits, listed in Table~\ref{tab:numbers_1} for the 92 action-angle groups, now put us in a position to address the basic questions we have about the properties of these action-angle groups. Table~\ref{tab:numbers_1} includes the group number of each group, the number of member stars N$_{*}$, the median action and angle coordinates, the best age estimate log $\tau$, the intrinsic age spread estimate $\sigma_{\ltau}$, and the metallicity [Fe/H] of the likelihood isochrone fitting. Additionally, for the 62 groups we identified in the cluster catalog by \citet{cantatgaudin_2020} as described in Section~\ref{iso_illustration}, we list the name of the identified literature cluster. Among the listed groups we find established clusters such as the Pleiades (Melotte 22), the Hyades (Melotte 25), Praesepe (NGC 2632), and Coma Berenices (Melotte 111). We also identify the Meingast 1 (Pisces-Eridanus) stellar stream by crossmatching our sources with \citet{Meingast_2019b}. However, a direct comparison and assignment of our groups to literature cluster is difficult, because each literature work uses a different approach to identify and select member stars of clusters, which leads to a lot of variation in the member selections (e.g. some of our action-angle groups are overlapping with the same literature cluster identified in \citet{cantatgaudin_2020}). A more detailed look at these different member selections additionally shows that our method selects more extended and loose structures (associations, tidal tails, other mechanisms of dispersal) as compared to \citet{cantatgaudin_2020} and \citet{hunt_2023}, whose member selections focus on cluster cores mainly.


\startlongtable
\begin{deluxetable*}{ccccccccccccc}
\tablecaption{This table provides an overview of the main parameters of the 92 action-angle groups. We provide the group number, the number of member stars N$_*$, and the median action and angle coordinates of all groups. Additionally, we state the age $\ltau$, the intrinsic age spread $\sigma_{\ltau}$, and the metallicity [$Fe/H$] as determined by our likelihood isochrone fitting for each group. For the subset of 62 groups which we identified in the cluster catalog by \citet{cantatgaudin_2020}, we list the literature name in the last column, as well as for the Meingast 1 (Pisces-Eridanus) stellar stream that we identified from \citet{Meingast_2019b}. A slightly modified version of this "group center table" is available at the CDS.} \label{tab:numbers_1}
\tablewidth{0pt}
\tablehead{
\colhead{Group} & \colhead{N$_{*}$} & 
\colhead{J$_R$} & \colhead{J$_{\phi}$} & \colhead{J$_Z$} & 
\colhead{$\Theta_R$} & \colhead{$\Theta_{\phi}$} & \colhead{$\Theta_Z$} & 
\colhead{$\ltau$} & \colhead{$\sigma_{\ltau}$} & \colhead{[$Fe/H$]} & \colhead{Cluster}\\
 &  & (kpc km s$^{-2}$) & (kpc km s$^{-2}$) & (kpc km s$^{-2}$) & (rad) & (rad) & (rad) &  &  &  & 
}
\startdata
1 & 207 & 3.08 & 1675.39 & 0.28 & 3.62 & 0.02 & 0.16 & 7.60 & 0.10 & 0.10 & IC 2602 \\
2 & 128 & 9.33 & 1599.05 & 0.34 & 3.61 & 0.05 & 2.85 & 8.20 & 0.20 & 0.10 &   \\
3 & 1072 & 10.4 & 1765.7 & 0.37 & 1.71 & 6.18 & 2.79 & 8.80 & 0.10 & 0.10 & Stock 2 \\
4 & 136 & 17.92 & 1569.64 & 0.92 & 1.94 & 6.1 & 3.63 & 8.80 & 0.05 & 0.10 & Mamajek 4 \\
5 & 232 & 3.12 & 1675.5 & 0.77 & 2.99 & 0.03 & 1.07 & 8.10 & 0.20 & 0.10 & UBC 1 \\
6 & 1059 & 14.01 & 1713.86 & 0.25 & 1.7 & 6.13 & 6.05 & 8.60 & 0.45 & 0.10 & Melotte 25 \\
7 & 1621 & 2.47 & 1677.09 & 0.13 & 3.66 & 0.02 & 1.22 & 7.30 & 0.10 & 0.10 & UPK 640 \\
8 & 231 & 13.64 & 1758.37 & 0.71 & 1.37 & 6.19 & 6.24 & 9.10 & 0.05 & 0.10 & NGC 6991 \\
9 & 3527 & 4.14 & 1660.97 & 0.37 & 3.1 & 6.23 & 5.14 & 8.10 & 0.15 & 0.10 & NGC 2516 \\
10 & 165 & 8.2 & 1744.32 & 0.6 & 1.84 & 6.16 & 0.11 & 8.10 & 0.25 & 0.10 &   \\
11 & 147 & 7.46 & 1659.19 & 0.2 & 3.25 & 0.02 & 3.8 & 8.50 & 0.35 & 0.10 &   \\
12 & 899 & 7.23 & 1653.34 & 0.37 & 3.35 & 0.02 & 3.49 & 8.00 & 0.05 & 0.10 & Melotte 22 \\
13 & 568 & 0.98 & 1862.0 & 0.48 & 5.39 & 0.07 & 4.48 & 8.00 & 0.05 & 0.10 & NGC 1039 \\
14 & 205 & 14.93 & 1827.63 & 0.22 & 5.26 & 0.18 & 3.31 & 8.60 & 0.05 & 0.10 & NGC 7092 \\
15 & 663 & 2.5 & 1832.2 & 1.05 & 6.15 & 0.02 & 3.64 & 8.30 & 0.30 & 0.10 & Meingast 1 \\
16 & 397 & 17.09 & 1945.29 & 1.0 & 5.51 & 0.13 & 5.71 & 8.90 & 0.05 & 0.10 & NGC 1662 \\
17 & 119 & 4.67 & 1677.42 & 0.4 & 3.57 & 0.02 & 3.66 & 8.10 & 0.05 & 0.10 &   \\
18 & 262 & 2.26 & 1688.44 & 0.27 & 3.33 & 0.01 & 6.16 & 7.90 & 0.35 & 0.10 &   \\
19 & 283 & 18.65 & 1933.09 & 0.32 & 5.55 & 0.12 & 1.38 & 8.30 & 0.20 & 0.10 & ASCC 41 \\
20 & 1017 & 13.38 & 1725.27 & 0.56 & 1.74 & 6.12 & 1.58 & 8.50 & 0.45 & 0.10 & NGC 2632 \\
21 & 280 & 41.22 & 1645.85 & 2.08 & 4.85 & 0.29 & 3.8 & 9.30 & 0.05 & 0.10 & Ruprecht 147 \\
22 & 785 & 1.75 & 1845.02 & 1.62 & 4.69 & 6.27 & 3.74 & 8.30 & 0.05 & 0.10 & NGC 2287 \\
23 & 204 & 30.61 & 1594.95 & 0.96 & 1.47 & 6.06 & 3.14 & 9.00 & 0.05 & 0.10 & Ruprecht 145 \\
24 & 222 & 7.73 & 1718.35 & 0.35 & 2.28 & 6.13 & 0.29 & 8.90 & 0.05 & 0.10 & NGC 2527 \\
25 & 325 & 4.72 & 1643.77 & 0.24 & 2.04 & 6.23 & 1.52 & 8.80 & 0.05 & 0.10 & NGC 6633 \\
26 & 142 & 7.48 & 1637.95 & 0.27 & 3.62 & 0.01 & 3.66 & 7.90 & 0.05 & 0.10 & UPK 545 \\
27 & 177 & 4.07 & 1809.3 & 0.15 & 6.09 & 0.01 & 3.38 & 8.60 & 0.35 & 0.10 & Alessi 9 \\
28 & 110 & 6.58 & 1736.49 & 0.13 & 1.15 & 6.19 & 5.71 & 8.00 & 0.15 & 0.10 & ASCC 99 \\
29 & 797 & 1.92 & 1795.07 & 1.61 & 0.69 & 6.25 & 4.5 & 8.10 & 0.15 & 0.10 & Blanco 1 \\
30 & 2790 & 0.31 & 1705.56 & 0.58 & 3.37 & 6.23 & 0.27 & 8.30 & 0.25 & 0.10 & NGC 3532 \\
31 & 266 & 2.02 & 1672.29 & 0.69 & 3.84 & 0.02 & 6.06 & 8.00 & 0.10 & 0.10 & UPK 612 \\
32 & 724 & 11.58 & 1976.23 & 0.88 & 5.59 & 0.1 & 5.3 & 8.70 & 0.10 & 0.10 & NGC 1647 \\
33 & 1399 & 0.93 & 1688.94 & 0.48 & 3.02 & 0.03 & 1.71 & 7.50 & 0.10 & 0.10 & Stephenson 1 \\
34 & 107 & 12.35 & 1849.22 & 1.11 & 5.83 & 0.02 & 3.28 & 8.80 & 0.05 & 0.10 & Ruprecht 98 \\
35 & 359 & 3.87 & 1757.89 & 1.82 & 2.18 & 6.25 & 3.71 & 9.20 & 0.05 & 0.10 & NGC 752 \\
36 & 232 & 2.59 & 1814.2 & 0.74 & 5.62 & 0.04 & 0.91 & 8.80 & 0.05 & 0.10 & Melotte 111 \\
37 & 171 & 4.63 & 1634.79 & 0.43 & 3.06 & 0.02 & 0.66 & 8.10 & 0.20 & 0.10 &   \\
38 & 241 & 3.29 & 1790.91 & 1.3 & 0.6 & 6.25 & 4.77 & 8.00 & 0.05 & 0.10 &   \\
39 & 294 & 4.92 & 1708.36 & 0.62 & 2.4 & 6.23 & 0.89 & 8.50 & 0.35 & 0.10 &   \\
40 & 161 & 17.43 & 1912.98 & 1.65 & 5.64 & 0.13 & 1.51 & 8.80 & 0.05 & 0.10 &   \\
41 & 152 & 0.95 & 1790.31 & 0.61 & 5.51 & 0.04 & 0.97 & 8.60 & 0.30 & 0.10 &   \\
42 & 229 & 10.87 & 1917.12 & 0.13 & 6.25 & 6.25 & 5.16 & 8.90 & 0.05 & 0.10 & Alessi 3 \\
43 & 871 & 0.87 & 1696.0 & 0.24 & 3.02 & 6.22 & 2.93 & 7.70 & 0.15 & 0.10 & BH 99 \\
44 & 381 & 3.38 & 1803.64 & 0.86 & 0.06 & 0.06 & 3.75 & 8.00 & 0.05 & 0.10 & Alessi 12 \\
45 & 1077 & 5.67 & 1740.7 & 0.31 & 0.22 & 6.23 & 1.33 & 7.90 & 0.45 & 0.10 & NGC 6124 \\
46 & 155 & 3.02 & 1867.78 & 0.16 & 5.79 & 0.04 & 1.63 & 8.80 & 0.05 & 0.10 &   \\
47 & 127 & 2.09 & 1657.79 & 0.54 & 3.37 & 0.04 & 1.79 & 7.90 & 0.30 & 0.10 &   \\
48 & 128 & 2.08 & 1712.15 & 0.14 & 2.56 & 6.2 & 2.62 & 7.60 & 0.05 & 0.10 & Trumpler 10 \\
49 & 387 & 3.32 & 1851.33 & 0.99 & 1.52 & 6.21 & 1.49 & 8.70 & 0.05 & 0.10 & NGC 2281 \\
50 & 640 & 9.01 & 1646.3 & 0.2 & 3.52 & 0.05 & 3.63 & 8.10 & 0.20 & 0.10 & Melotte 22 \\
51 & 226 & 13.22 & 1963.6 & 0.25 & 5.56 & 0.12 & 2.5 & 8.10 & 0.05 & 0.10 & COIN-Gaia 13 \\
52 & 297 & 7.88 & 1855.43 & 0.28 & 5.72 & 0.06 & 6.04 & 8.80 & 0.05 & 0.10 &   \\
53 & 327 & 2.47 & 1772.2 & 0.02 & 0.29 & 6.26 & 6.16 & 8.30 & 0.20 & 0.10 & NGC 6475 \\
54 & 462 & 7.97 & 1974.53 & 0.78 & 5.74 & 0.07 & 5.08 & 8.70 & 0.15 & 0.10 & NGC 1647 \\
55 & 163 & 17.34 & 1917.93 & 0.35 & 5.53 & 0.11 & 1.26 & 8.40 & 0.30 & 0.10 & ASCC 41 \\
56 & 122 & 3.26 & 1845.88 & 0.19 & 0.68 & 6.25 & 1.23 & 8.70 & 0.35 & 0.10 & Platais 3 \\
57 & 171 & 1.28 & 1733.91 & 0.13 & 2.35 & 6.2 & 2.6 & 7.70 & 0.05 & 0.10 & Trumpler 10 \\
58 & 717 & 1.5 & 1723.34 & 0.12 & 2.44 & 6.2 & 2.65 & 7.60 & 0.05 & 0.10 & Trumpler 10 \\
59 & 475 & 4.44 & 1605.23 & 0.21 & 2.45 & 6.26 & 1.96 & 8.90 & 0.05 & 0.00 & IC 4756 \\
60 & 793 & 12.78 & 1937.18 & 0.38 & 5.95 & 0.07 & 2.18 & 9.00 & 0.20 & 0.10 &   \\
61 & 243 & 14.54 & 1851.62 & 0.83 & 6.13 & 0.09 & 0.93 & 9.00 & 0.05 & 0.10 & Alessi 62 \\
62 & 220 & 16.2 & 1561.3 & 1.04 & 2.05 & 6.12 & 3.59 & 7.90 & 0.05 & 0.10 & Mamajek 4 \\
63 & 428 & 4.07 & 1744.38 & 0.17 & 2.41 & 6.18 & 2.29 & 8.00 & 0.05 & 0.10 & NGC 2422 \\
64 & 653 & 2.47 & 1765.78 & 0.17 & 2.39 & 6.19 & 2.31 & 8.00 & 0.05 & 0.10 & NGC 2422 \\
65 & 121 & 7.48 & 1731.42 & 0.15 & 1.22 & 6.17 & 5.72 & 8.10 & 0.20 & 0.10 &   \\
66 & 114 & 4.62 & 1653.71 & 0.26 & 4.92 & 0.08 & 0.15 & 8.90 & 0.05 & 0.10 & ASCC 90 \\
67 & 208 & 16.46 & 1886.53 & 0.52 & 5.65 & 0.11 & 2.3 & 8.90 & 0.15 & 0.10 &   \\
68 & 509 & 0.1 & 1715.53 & 0.55 & 3.04 & 6.22 & 0.26 & 8.60 & 0.05 & 0.10 & NGC 3532 \\
69 & 107 & 2.17 & 1794.33 & 0.78 & 0.27 & 6.26 & 4.21 & 8.40 & 0.25 & 0.10 &   \\
70 & 875 & 7.83 & 1899.92 & 0.37 & 5.65 & 0.09 & 2.06 & 8.90 & 0.10 & 0.10 &   \\
71 & 179 & 4.24 & 1673.15 & 0.54 & 3.49 & 0.03 & 1.62 & 9.30 & 0.20 & 0.10 &   \\
72 & 207 & 3.45 & 1717.36 & 0.78 & 2.14 & 6.19 & 1.11 & 8.60 & 0.40 & 0.10 &   \\
73 & 2677 & 16.3 & 1914.66 & 0.35 & 5.74 & 0.1 & 2.34 & 9.00 & 0.20 & 0.10 &   \\
74 & 159 & 1.65 & 1719.6 & 0.14 & 2.49 & 6.2 & 2.76 & 7.70 & 0.20 & 0.10 &   \\
75 & 477 & 13.93 & 1720.57 & 0.37 & 1.59 & 6.11 & 1.14 & 9.10 & 0.20 & 0.10 &   \\
76 & 1176 & 0.84 & 1822.89 & 0.19 & 6.2 & 6.25 & 3.67 & 7.60 & 0.10 & 0.10 & Collinder 135 \\
77 & 287 & 7.68 & 1575.09 & 0.3 & 4.34 & 0.05 & 5.72 & 8.40 & 0.05 & 0.10 & NGC 6025 \\
78 & 162 & 21.5 & 1872.02 & 0.38 & 5.56 & 0.16 & 6.1 & 8.90 & 0.15 & 0.10 &   \\
79 & 536 & 8.05 & 1950.77 & 2.1 & 5.61 & 0.02 & 1.2 & 8.70 & 0.05 & 0.10 & NGC 2548 \\
80 & 624 & 16.45 & 1879.6 & 0.06 & 5.57 & 0.12 & 2.0 & 9.10 & 0.25 & 0.10 &   \\
81 & 5236 & 2.7 & 1684.49 & 0.22 & 2.78 & 6.26 & 0.83 & 8.10 & 0.45 & 0.10 & Platais 3 \\
82 & 208 & 4.17 & 1642.18 & 0.44 & 3.03 & 6.27 & 5.3 & 8.70 & 0.40 & 0.10 &   \\
83 & 137 & 6.28 & 1618.91 & 0.39 & 3.0 & 0.01 & 0.69 & 8.10 & 0.15 & 0.10 &   \\
84 & 704 & 2.87 & 1868.11 & 1.5 & 5.02 & 0.01 & 3.8 & 8.30 & 0.05 & 0.10 & NGC 2287 \\
85 & 213 & 3.02 & 1824.24 & 1.85 & 0.26 & 6.22 & 5.08 & 8.90 & 0.05 & 0.10 & NGC 1901 \\
86 & 161 & 17.58 & 1940.0 & 0.97 & 5.97 & 0.05 & 3.87 & 8.90 & 0.05 & 0.10 &   \\
87 & 575 & 0.1 & 1744.3 & 0.6 & 6.17 & 6.23 & 0.25 & 8.10 & 0.05 & 0.10 & NGC 3532 \\
88 & 470 & 1.22 & 1661.24 & 0.21 & 4.86 & 0.03 & 0.64 & 8.50 & 0.10 & 0.10 & NGC 6281 \\
89 & 488 & 3.88 & 1919.99 & 1.14 & 5.34 & 0.1 & 5.47 & 8.90 & 0.25 & 0.10 & NGC 1342 \\
90 & 254 & 2.63 & 1704.19 & 0.03 & 2.34 & 6.28 & 1.0 & 7.80 & 0.05 & 0.10 & NGC 7058 \\
91 & 122 & 18.47 & 1906.46 & 0.45 & 5.8 & 0.06 & 1.1 & 8.70 & 0.05 & 0.10 &   \\
92 & 111 & 4.04 & 1619.59 & 0.55 & 4.1 & 0.04 & 4.3 & 7.80 & 0.05 & 0.10 & Alessi 24 \\
\enddata
\end{deluxetable*}


\subsubsection{How old are the groups? Are they mono-age?}

Figure~\ref{fig:values} shows the distribution of ages (log $\tau$) among the 92 groups, which range from 20~Myrs to 2~Gyrs: all are much younger than the mean age of Galactic disk stars ($\sim 3-5$~Gyrs). For our purpose, the definitions of ``young'' {\it vs.} ``old'' revolve around dynamical time-scales: one central timescale, the orbital period at the Solar radius is indicated by the blue dotted vertical line, for which we adopt $\mathrm{log}\tau = 8.33$.  We find that half (43) of the groups are young in this sense, younger than one dynamical period. The 49 older groups have ages $\tau=1-7~t_{orbit}\ll t_{Hubble}$.

Our fitting also shows that almost 80$\%$ (72) of the groups seem to be "effectively mono-age", with $\sigma_{\ltau} \leq$ 0.2; only 20 groups show a modest age dispersion of $\sigma_\ltau >$ 0.2. The value of 0.2 to distinguish these two regimes was chosen by comparing $\sigma_\ltau$ to the uncertainty $\delta_{\ltau}$ of the isochrone fit, which is typically around 0.1 (see Section~\ref{dispersion}). The intrinsic age spread is indicated in the color coding of the bars in Figure~\ref{fig:values}, where groups with significant intrinsic age spread are shown in magenta, and mono-age groups are shown in turquoise. 

This result immediately shows that none of these groups are just stars herded together by orbital resonances, as a wide intrinsic age spread would be expected in this case.

The best fitting metallicities fall almost exclusively on $[Fe/H]= 0.1$, with only one group showing a metallicity of $[Fe/H]= 0.0$. They are essentially all Solar [Fe/H], as expected for young stars near the Solar radius. Whether the slightly super-Solar values, surprising in light of the [Fe/H] of the present-day ISM at the Solar radius, reflect slight [M/H]-systematics in the isochrone fitting, remains open. Our metallicities are purely photometric, and draw in many cases heavily on the colors of lower main-sequence objects, whose isochrone-predicted colors are known to be uncertain. Therefore, our metallicities may well be subject to systematic errors at the 0.2 dex level. We consider this to be not relevant for our analysis, as the covariance between ages and metallicities are not a source of serious systematics in the age determinations.

\begin{figure*}[ht!]
\includegraphics[width=\textwidth]{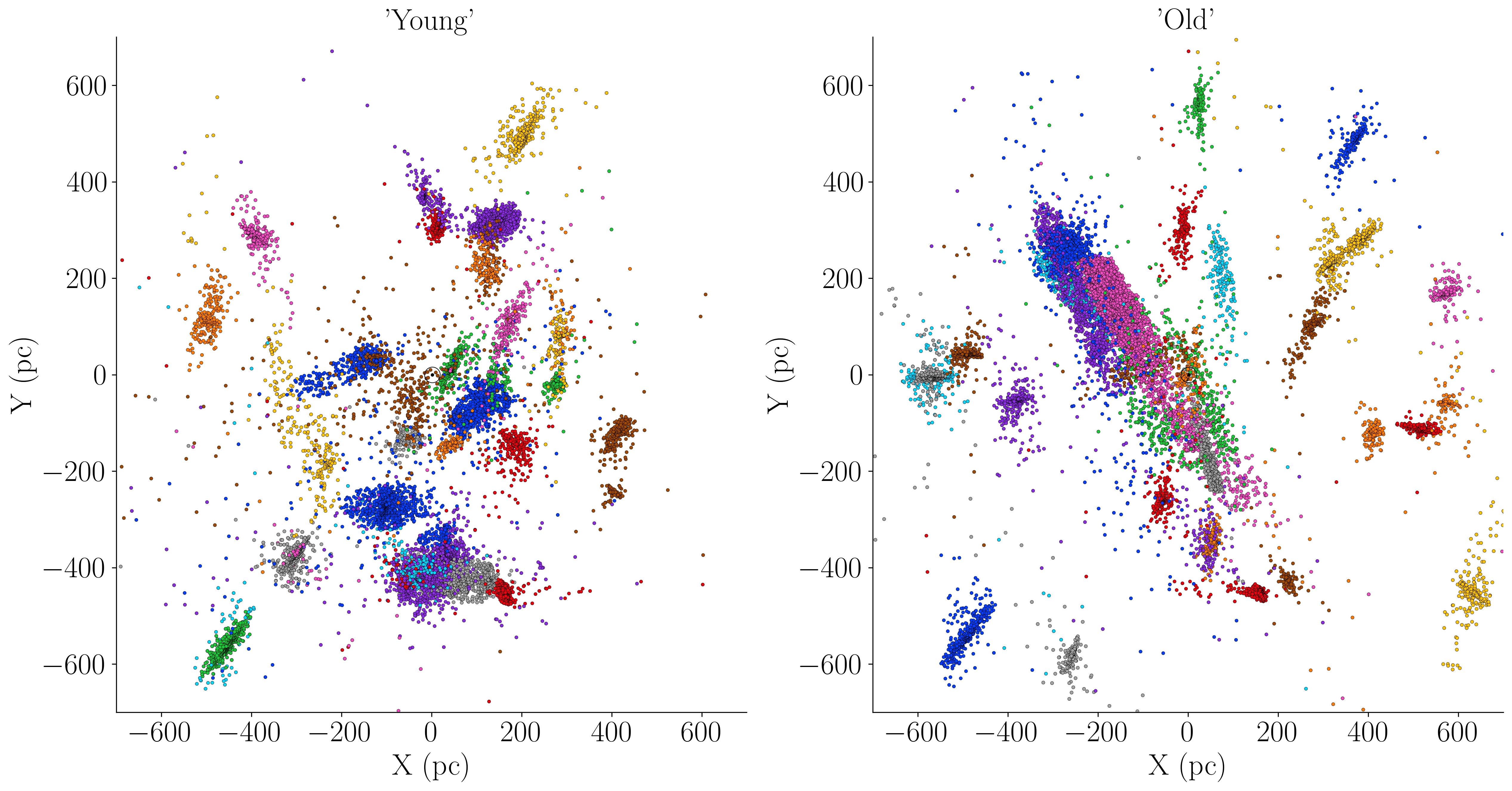}
\caption{Top-down views on the 92 action-angle groups in Galactic Cartesian coordinates, where the Sun is located at (X,Y) = (0,0). Left: 43 young groups with ages $<$ 250 Myrs. Right: 49 old groups with ages $>$ 250 Myrs. Their morphologies are either heterogeneous, or show a tight clustered ``core'' with or without tidal (or uncertainty-driven) tails.} \label{fig:XY}
\end{figure*}

\subsubsection{Spatial Structure of the Groups}\label{sec:groupstructure}
The spatial extent of the 92 groups (color-coded) is shown in Figure~\ref{fig:XY} in a top-down view onto the Milky Way.  For a better overview, the left panel shows the 43 groups younger than one Galactic revolution ($\tau <$ 250 Mrys), and the right panel shows the 49 old groups with ages $>$ 250 Myrs. This figure reveals a variety of spatial distributions and the internal spatial extents of the groups: some heterogeneous morphologies with little center-to-edge structure, and often an extent of $\gtrsim 100$~pc; some show a tight clustered ``core'' with or without tidal (or uncertainty-driven) tails. For the most part, these tails extend along the direction of the Galactic rotation as extended for tidal tails, as found and discussed previously in various papers \citep{meingast_2019, roeser_2019, roeser_2019b, meingast_2021, fuernkranz_2019, tang_2019}. Overall, most groups extend across several hundred parsecs on the sky.

\subsubsection{Overlap with Existing Cluster Catalogs}

As was already apparent from the age comparison with \citet{cantatgaudin_2020}, many of our groups are in published catalogs.
Specifically,  33\% of stars that are in one of our groups have a crossmatch with the \citet{cantatgaudin_2020} cluster catalog,  
52\% with the \citet{hunt_2023} catalog, and 49\% with the catalog of \citet{kounkel_2019}, known to be complete by also contaminated \citep{zucker_2022}.

Among our 92 groups, 22 have no stars in common with any clusters in \citet{cantatgaudin_2020}; the majority of our groups (51) have stars that overlap with exactly one cluster in \citet{cantatgaudin_2020}. Remarkably, for 19 of our groups the stars within one group overlap with several (up to ten) clusters in the \citet{cantatgaudin_2020} catalog. This shows that our algorithm -- designed to also identify loose orbit-space groups -- can and does encompass several clusters if they are on near-identical orbits at very similar orbital phases.  This is very much in line with the extended structure of our groups, already discussed in Section~\ref{sec:groupstructure}.

\section{Summary and Discussion} \label{sec:summary}

We have set out to devise and implement a new stringent two-step method to determine isochrone ages of action-angle groups in the nearby Galactic disk, even if they are sparse and widely dispersed. First, we overcome the limitation that the initial sets of stars, found to be clustered in 6D action-angle space, have a limited T$_{eff}$-range that hinders age dating with isochrones. We do this by projecting the 6D action-angle patch into the 5D space of position and proper motion, where the Gaia catalog is far more extensive. We can identify typically an order of magnitude more likely members. 
The second step is the actual isochrone fitting, as a function of both age and metallicity. This fitting accounts for a number of factors important here: the group members may be widely distributed in sky position, distance and reddening; they may have sizeable parallax uncertainties; and we must account for binaries and other ``outliers``, such as stars that are not group members. We do not only determine the mean age of each group, but we also test whether any group's CMD provides evidence for a significant intrinsic age spread.
We find that our approach provides well-constrained ages, age dispersions and metallicities. 

 We have found that all of the identified 92 action-angle groups are ``young`` in a global Galaxy formation sense: $\tau \ll \tau_{disk}\approx 8-11$~Gyrs, as they all have $\tau \le 2$~Gyrs. About half of them are even younger than a dynamical period ($<t_{orbit}\approx$ 250 Myrs). The large majority of them is consistent with a mono-age population, $\sigma_{\ltau} \leq$ 0.2~mag. 
 
 Not accounting for differential selection effects, we find that the distribution $p\bigl(\ltau\bigr)$ is broadly constant between $\ltau=7$ and 9, which would naively imply that $p(\tau)\propto \tau^{-1}$. 

We also find that some groups are spatially extended and most likely entirely unbound, while others are spatially compact.

 These findings fit together in a picture in which these action-angle groups indeed reflect ensembles of stars born at nearly the same time in nearly the same place, and are now dispersing to become field stars.   How rapidly these groups of stars disperse, and how compact they once were, would require quantitative modelling of their actions and angles, which we will carry out in a forthcoming paper.

 We have also found that we are not severely limited in the age determination by the group member sample size with our approach, as our algorithm provides almost the exact same age estimates for all groups if we randomly sample only half of their stars. This bodes well for applying the same approach in a forthcoming paper to the smallest and most diffuse groups we can identify, such as the \emph{pearls on a string} found by \citet{coronado2022} for a set of action angle groups: these are -- often numerous -- further clumps in angle (or orbital phase) space on nearly the same orbit as the initially identified action-angle group.
 

\section*{Acknowledgements}
We wish to thank the anonymous referee for his/her useful comments, which improved both the clarity and quality of this study. We also thank the MPIA Milky Way group and Andy Gould for providing useful comments. V.F. acknowledges support from the International Max Planck Research School for Astronomy and
Cosmic Physics at Heidelberg University (IMPRS-HD). This work has made use of data from the European Space Agency (ESA) mission
{\it Gaia} (\url{https://www.cosmos.esa.int/gaia}), processed by the {\it Gaia}
Data Processing and Analysis Consortium (DPAC,
\url{https://www.cosmos.esa.int/web/gaia/dpac/consortium}). Funding for the DPAC
has been provided by national institutions, in particular the institutions
participating in the {\it Gaia} Multilateral Agreement.

\software{We acknowledge various Python packages that were used in the data analysis of this work, including astropy \citep{2013A&A...558A..33A,2018AJ....156..123A}, NumPy \citep{numpy}, SciPy \citep{scipy}, Matplotlib \citep{matplotlib}, and scikit-learn \citep{scikit-learn}. This research also mady use of the TORUSMAPPER package \citep{binney_2016}, which is part of the AGAMA library \citep{vasiliev_2019}. 
          }



\bibliography{sample631}{}
\bibliographystyle{aasjournal}

\end{document}